# Key Encapsulation Mechanism-Based Integrated Encryption Scheme (KEM-IES)

Abel C. H. Chen
*Information & Communications Security Laboratory, Chunghwa Telecom Laboratories*
Taoyuan, Taiwan
Email: chchen.scholar@gmail.com; ORCID: 0000-0003-3628-3033

***Abstract*—The Elliptic Curve Integrated Encryption Scheme (ECIES) is widely regarded as a practical method and has been adopted by multiple standards. However, the advancement of quantum computing technologies poses potential security risks to ECIES. Therefore, this study proposes a Key Encapsulation Mechanism-Based Integrated Encryption Scheme (KEM-IES), which enhances resistance to quantum attacks by incorporating a Post-Quantum Cryptography (PQC)-based Key Encapsulation Mechanism (KEM). Furthermore, the study integrates the Ascon algorithm, released by the National Institute of Standards and Technology (NIST) in August 2025, to further improve computational efficiency and enable applicability to embedded systems. The proposed KEM-IES and its Ascon-based variant are implemented on a Raspberry Pi 4, and evaluations are conducted to compare the performance of ECIES and KEM-IES.**

## I. INTRODUCTION

Quantum computing technologies have continued to advance, bringing increased attention to the issue of "Harvest Now, Decrypt Later" (HNDL) [1]. The Elliptic Curve Integrated Encryption Scheme (ECIES) is one of the mainstream encryption mechanisms in current use [2], and a number of standards (e.g., IEEE 1609.2 [3], IEEE 1609.2.1 [4], and IETF RFC 8446 [5]) still rely on ECIES and Elliptic Curve Diffie-Hellman (ECDH) for encryption and key exchange. However, quantum computing has the potential to compromise the security of the Elliptic Curve Discrete Logarithm Problem (ECDLP) and Elliptic Curve Cryptography (ECC) [6]. As a result, both ECIES and ECDH lack quantum resistance and face increasing vulnerability under HNDL-style threats.

Given these concerns, the design of an Integrated Encryption Scheme (IES) with quantum-resistant properties has become an urgent challenge. Therefore, this study proposes a Key Encapsulation Mechanism-Based IES (KEM-IES) that incorporates Module Lattice-Based Key Encapsulation Mechanisms (ML-KEM) [7] or Hamming Quasi-Cyclic (HQC) [8], both of which are Post-Quantum Cryptography (PQC) algorithms selected by the National Institute of Standards and Technology (NIST). These mechanisms provide the foundation for achieving quantum security. Furthermore, the study integrates the lightweight cryptography algorithm (i.e., Ascon algorithm [9]) to further enhance computational efficiency. The proposed design is implemented and evaluated on a Raspberry Pi 4 to validate its effectiveness through empirical analysis. The main contributions of this study are summarized as follows.

- Quantum Security: The proposed KEM-IES employs a ML-KEM during the key exchange phase. Its underlying mathematical hardness relies on the Shortest Vector Problem (SVP) and the Learning With Errors (LWE) problem, thereby providing resistance against quantum computational attacks.
- Lightweight Cryptography: The Ascon-based KEM-IES variant proposed in this study enhances encryption performance. By utilizing Ascon in both the key derivation and encryption phases, the scheme achieves improved computational efficiency.
- Empirical Evaluation: To assess performance on embedded systems, the study implements the proposed KEM-IES and the Ascon-based KEM-IES on a Raspberry Pi 4. Their performance is practically compared with that of ECIES.

## II. RELATED WORK

Section II.A explains the flows of the ECIES, while Section II.B describes how ECIES is applied in V2X communications and outlines the structure of the encrypted SPDU.

### *A. Elliptic Curve Integrated Encryption Scheme*

In ECIES, the process begins with an ECDH key exchange to establish a Key-Encryption Key (KEK), denoted as $z_1$. As illustrated in Fig. 1, assume that Bob's ECC public key $v_b$ is publicly available and known to Alice. Alice then generates an ephemeral ECC key pair (e.g., a NIST P-256 key pair), consisting of her private key $u_a$ and corresponding public key $v_a$. Alice transmits $v_a$ to Bob, after which both parties compute a shared elliptic-curve point by multiplying their private key with the other party's public key. The x-coordinate of this shared point is extracted and used as $z_1$ the result of the key exchange.

The KEK $z_1$, along with the shared contextual information $p$ (e.g., the hash value of the request message for integrity checking), is then passed into a Key Derivation Function (KDF). In this scheme, KDF2 based on the Secure Hash Algorithm 256 (SHA-256) is used to derive a 256-bit value. The first 128 bits of the output are taken as $k_1$, and the remaining 128 bits as $k_2$. Alice subsequently generates an AES-128–based Data-Encryption Key (DEK), denoted as $s$, and uses $k_1$ to encrypt $s$, producing the encrypted DEK $c$, which is then sent to Bob. In the IEEE 1609.2 [3] and IEEE 1609.2.1 [4] standards, the encryption of $s$ is performed using $c = s \oplus k_1$. Therefore, once Bob obtains $z_1$ and completes the KDF computation, he can recover the DEK by computing $s = c \oplus k_1$. Finally, the encrypted DEK $c$ and $k_2$ are input into a keyed-Hash Message Authentication Code (HMAC) function to generate the Message Authentication Code (MAC) $t$, which serves as a tag. Alice transmits $t$ to Bob, who recomputes the MAC using $c$ and $k_2$ and verifies whether it matches the received tag. Successful verification confirms message integrity, allowing both parties to use the DEK $s$ with AES-128 for subsequent data encryption and decryption.

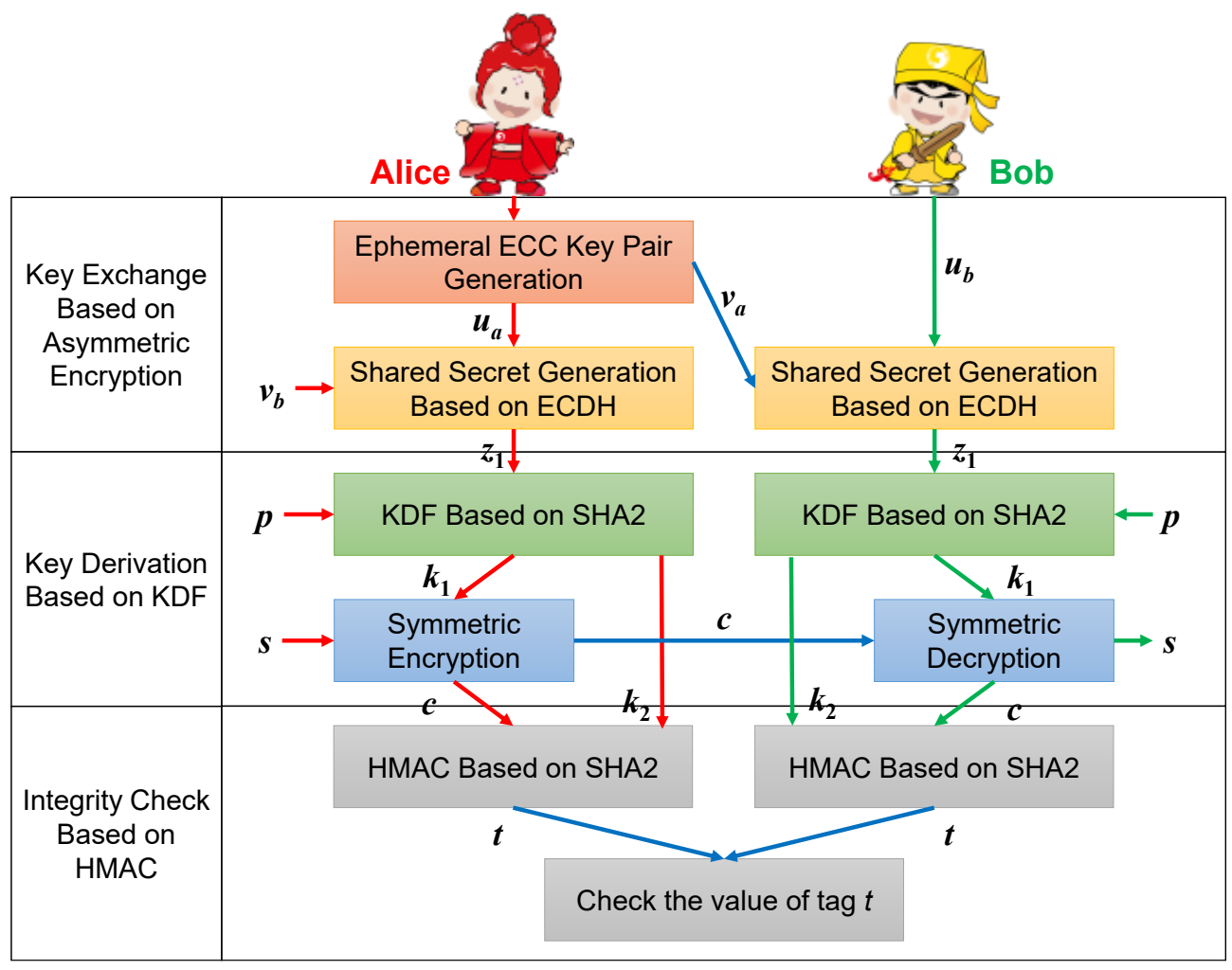

**Fig. 1.** The flows of ECIES.

### *B. The Encrypted Secure Protocol Data Unit in IEEE 1609.2*

As described in Section II.A, the ECIES process requires Alice to transmit three items to Bob: Alice's ECC public key $v_a$ (which functions as the encrypted KEK), the encrypted DEK $c$, and the tag $t$. Accordingly, IEEE 1609.2 [3] specifies the EncryptedDataEncryptionKey structure to encapsulate these elements. Furthermore, the actual application data to be transmitted is encrypted using the DEK $s$ as the AES-128 key, together with a randomly generated nonce. The data are encrypted under the AES-128 Cipher Block Chaining (CBC)-MAC Mode (CCM), producing the CCM ciphertext. IEEE 1609.2 [3] defines the Ciphertext structure to store both the nonce and the resulting CCM ciphertext. Finally, the EncryptedDataEncryptionKey structure and the Ciphertext structure are combined to form the encrypted SPDU, as illustrated in Fig. 2.

## III. The Proposed Methods

Section III.A explains the flows of the proposed KEM-IES. Section III.B then integrates ECIES and KEM-IES to construct a hybrid IES. Finally, Section III.C describes how the proposed schemes are applied to V2X communications and presents the revised encrypted SPDU structure.

### *A. Key Encapsulation Mechanism-Based Integrated Encryption Scheme*

In the KEM-IES, the process begins with establishing the KEK $z_2$ using a PQC-based Key Encapsulation Mechanism (KEM). As illustrated in Fig. 3, assume that Bob's KEM public key $K_{b,pub}$ is publicly available and held by Alice. Using $K_{b,pub}$, Alice executes the PQC-based KEM to generate the shared secret $z_2$ (i.e., the KEK) and the encapsulated key value $e$ (i.e., the encrypted KEK). Alice then sends $e$ to Bob. Bob applies his KEM private key $K_{b,priv}$ to decapsulate $e$ to obtain the KEK $z_2$.

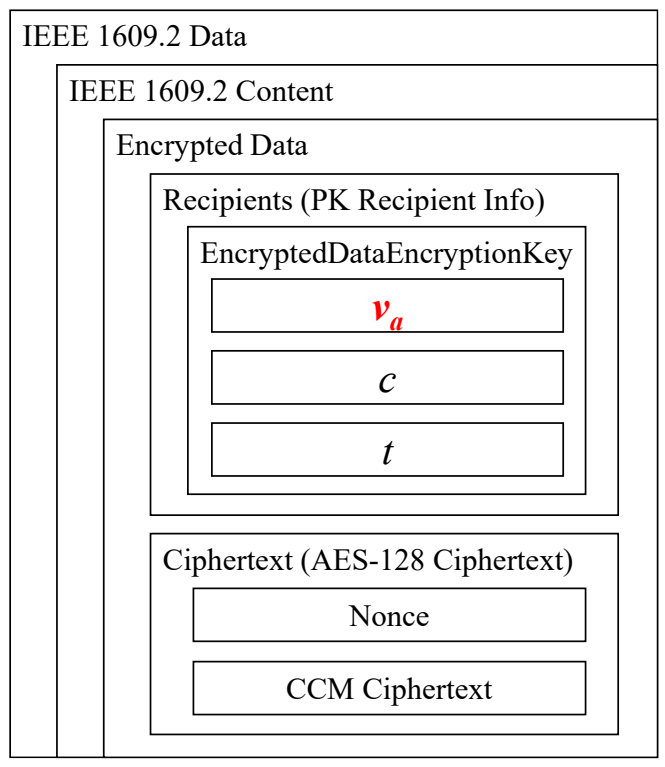

**Fig. 2.** The structure of encrypted SPDU in IEEE 1609.2 [3].

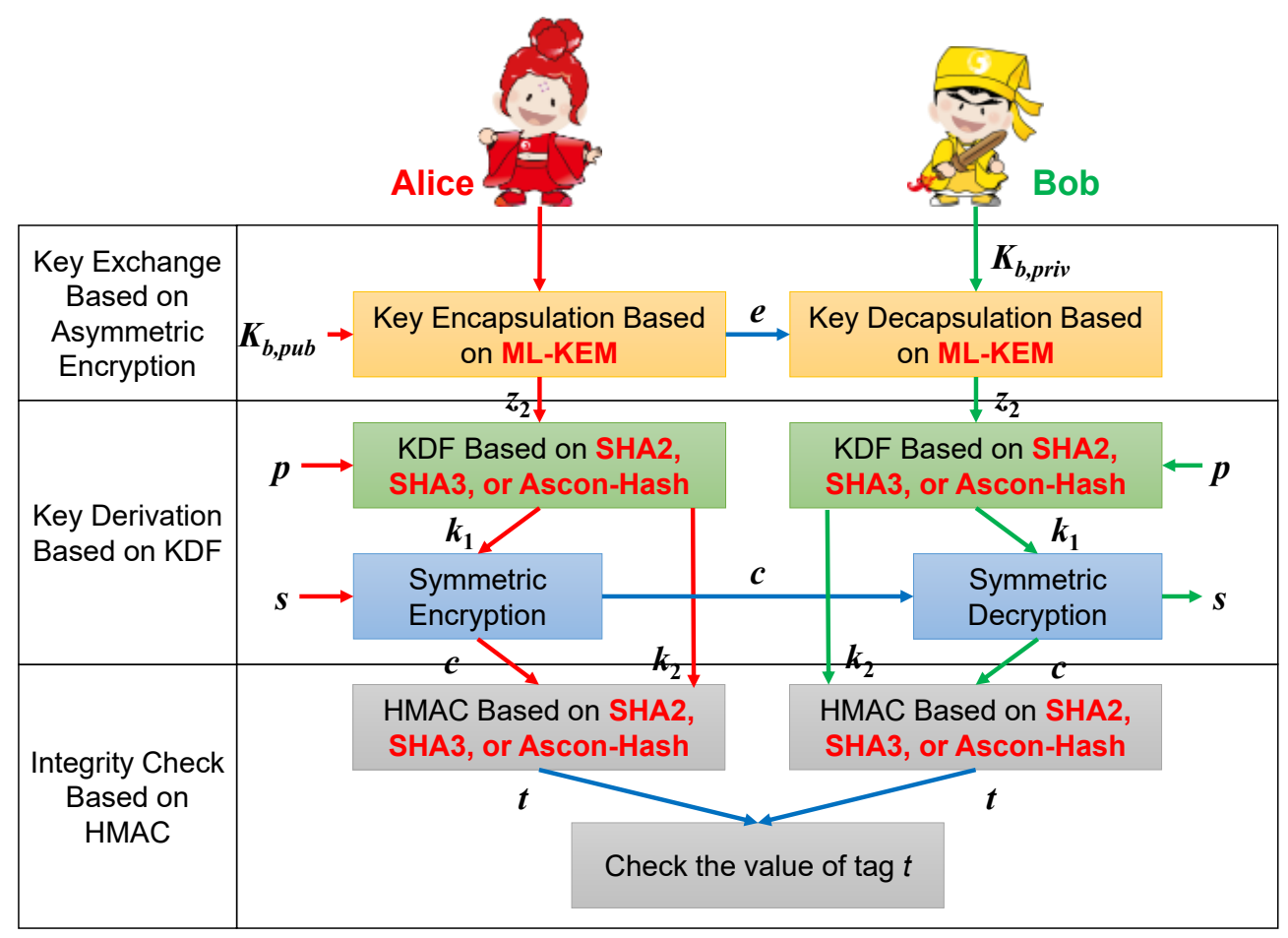

**Fig. 3.** The flows of KEM-IES.

Modifications can be made to the key derivation and integrity checking processes based on ECIES. In ECIES, the KDF is constructed using SHA-2; this may be replaced with SHA-3 or Ascon-Hash. After inputting the KEK $z_2$ together with the shared contextual information $p$, the derived outputs $k_1$ and $k_2$ are obtained. Alice then generates a DEK $s$, where the symmetric encryption algorithm may be AES or Ascon-128 Authenticated Encryption with Associated Data (AEAD). Using $k_1$, Alice encrypts $s$ to form the encrypted DEK $c$, which is subsequently transmitted to Bob. For integrity checking, the hash function used in the HMAC computation may likewise be replaced with SHA-3 or Ascon-Hash. The tag $t$ is then produced by applying the HMAC function to the encrypted DEK $c$ together with $k_2$.

### *B. Hybrid Integrated Encryption Scheme*

In the proposed hybrid IES, Alice is assumed to possess both Bob's ECC public key $v_b$ and Bob's KEM public key $K_{b,pub}$. She executes the ECDH key exchange described in Section II.A, as well as the PQC-based KEM described in Section III.A, thereby generating her ECC public key $v_a$ and the encapsulated key value $e$. These values are transmitted to Bob. Both parties can then compute the results $z_1$ and $z_2$, as illustrated in Fig. 4.

In the key derivation process, three inputs are provided to the KDF: $z_1$, $z_2$, and the shared contextual information $p$. The KDF then produces the derived keys $k_1$ and $k_2$. While the original KDF is based on SHA-2, it may be replaced with SHA-3 or Ascon-Hash. Subsequently, $k_1$ is used to encrypt the DEK $s$, producing the encrypted DEK $c$, which is transmitted to Bob. For integrity checking, the hash function in the HMAC computation can similarly be replaced with SHA-3 or Ascon-Hash, and the tag $t$ is calculated by applying the HMAC to the encrypted DEK $c$ together with $k_2$.

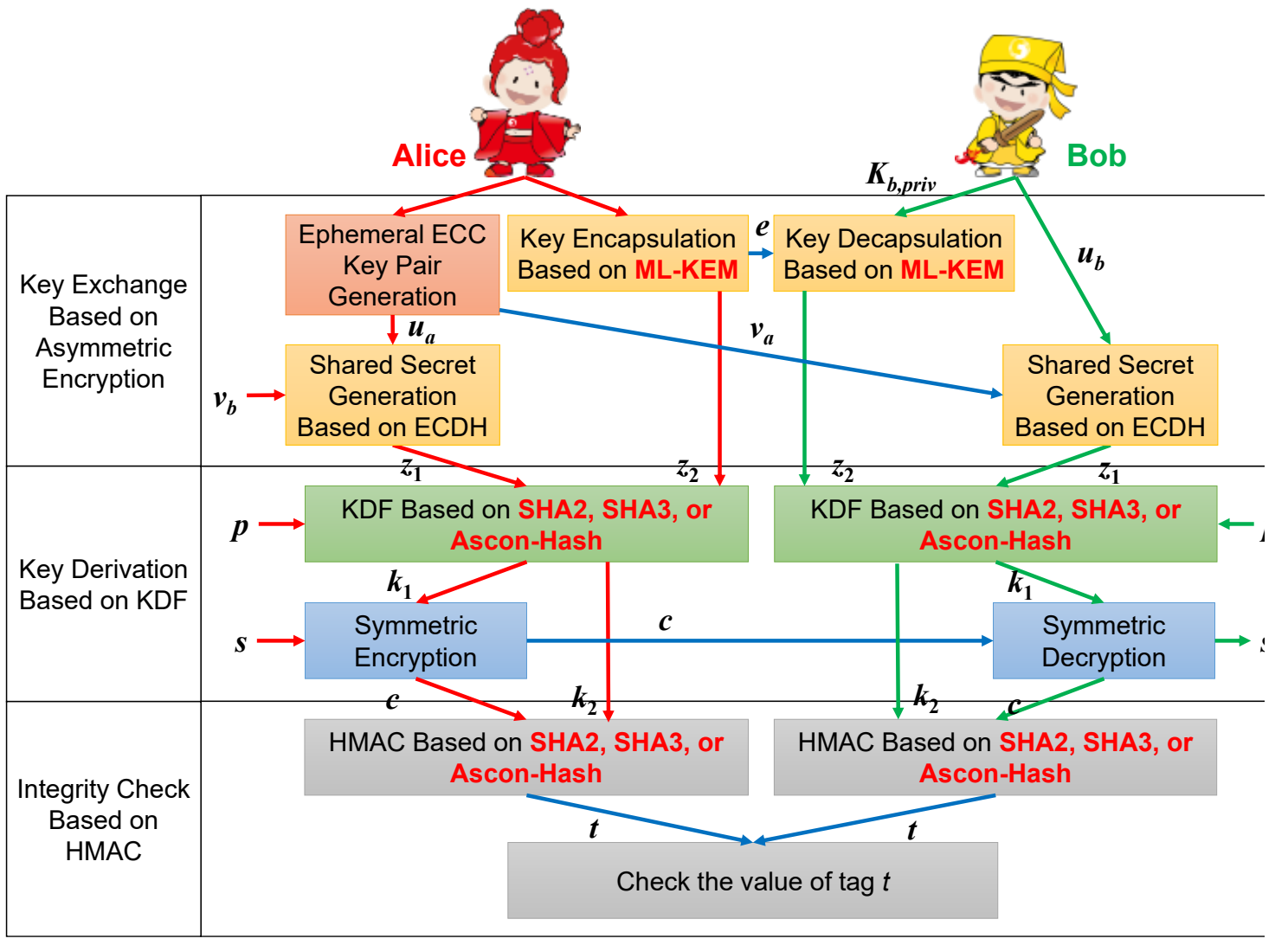

**Fig. 4.** The flows of Hybrid-IES.

### C. The Revised Encrypted SPDU

In the KEM-IES, Alice transmits three items to Bob: the encapsulated key value $e$ (e.g., the encrypted KEK), the encrypted DEK $c$, and the tag $t$. In the hybrid IES, Alice transmits four items: her ECC public key $v_a$ (i.e., part of the encrypted KEK), the encapsulated key value $e$ (i.e., the other part of the encrypted KEK), the encrypted DEK $c$, and the tag $t$. Accordingly, the EncryptedDataEncryptionKey structure can be revised to accommodate the encrypted KEK designed in this study: in KEM-IES, the encrypted KEK consists of $e$, whereas in the hybrid IES, the encrypted KEK consists of $e \| v_a$, as illustrated in Fig. 5. The formats of the encrypted DEK $c$ and the tag $t$ remain unchanged. Similarly, the Ciphertext structure retains the same formats for the nonce and CCM ciphertext fields, although the symmetric encryption algorithm may be either AES or Ascon.

## IV. Practical Experimental Results and Discussions

### A. Practical Experimental Environment

To evaluate and compare the ECIES, the KEM-IES, and the hybrid IES, implementations were conducted on a Raspberry Pi 4. In these implementations, ECIES utilized the ECC P-256 curve, while KEM-IES employed ML-KEM-512 and HQC-128 for PQC-based KEM. For hash computations, SHA-256, SHA3-256, SHAKE-128, and Ascon-Hash256 were implemented and compared. Furthermore, symmetric encryption performance was evaluated using both AES-128 and Ascon-AEAD128.

### B. Computation Time Comparison

This study compares the key generation time (Fig. 6), KEK encryption time (Fig. 7), DEK encryption time (Fig. 8), and data encryption time (Fig. 9), with all measurements reported in milliseconds. ML-KEM demonstrates the highest efficiency in key pair generation and key encapsulation, resulting in significantly lower computation times compared with other methods, as observed in Fig. 6, Fig. 7, and Fig. 8. In contrast, HQC exhibits the lowest efficiency in key pair generation and key encapsulation, leading to notably higher computation times than the other methods. ECIES, which is based on ECC, achieves computational efficiency between that of ML-KEM and HQC. In the hybrid IES, both the ECDH key exchange and the PQC-based KEM are executed, resulting in computation times approximately equal to the sum of those for ECIES and KEM-IES.

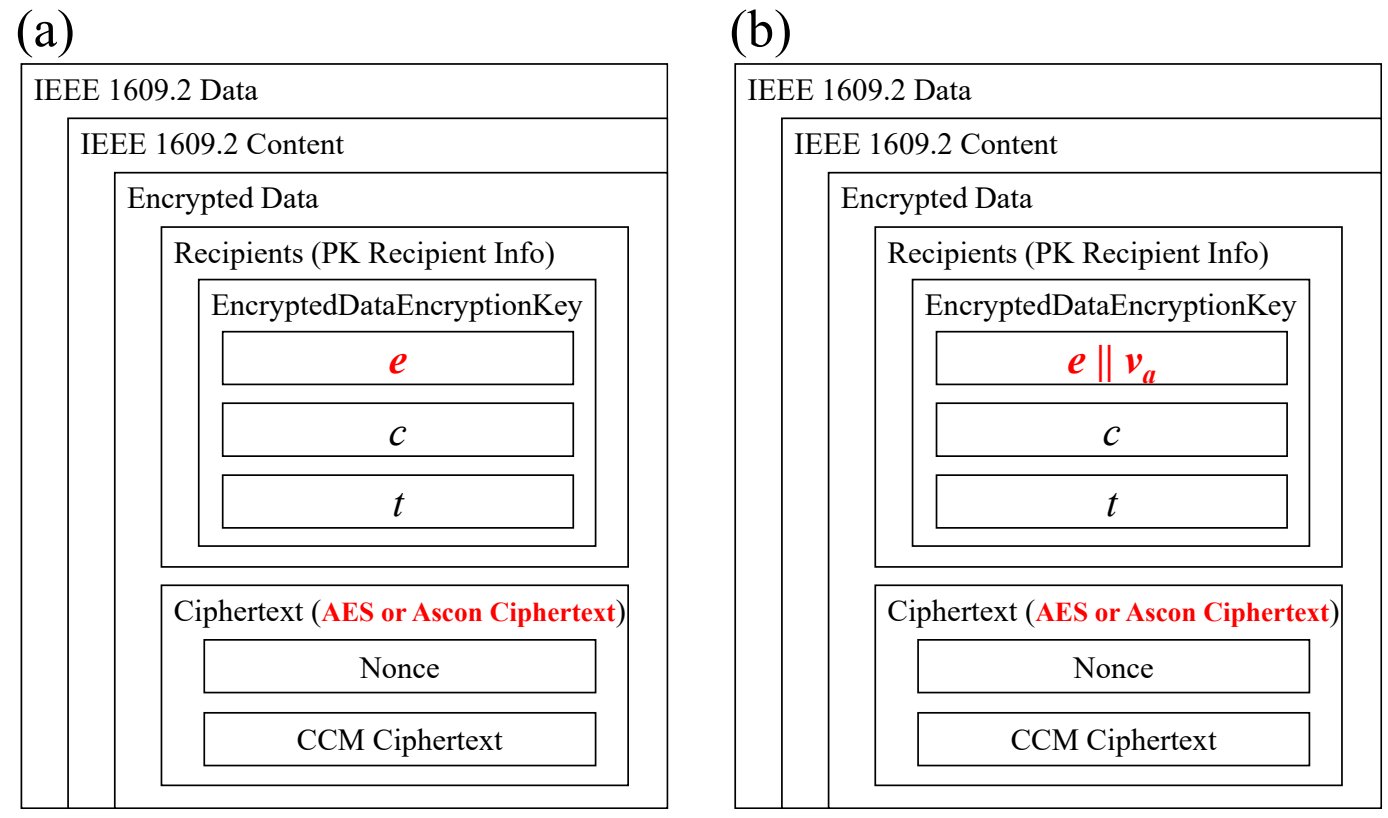

**Fig. 5.** Revised encrypted SPDU structures based on IEEE 1609.2: (a) for KEM-IES and (b) for Hybrid IES.

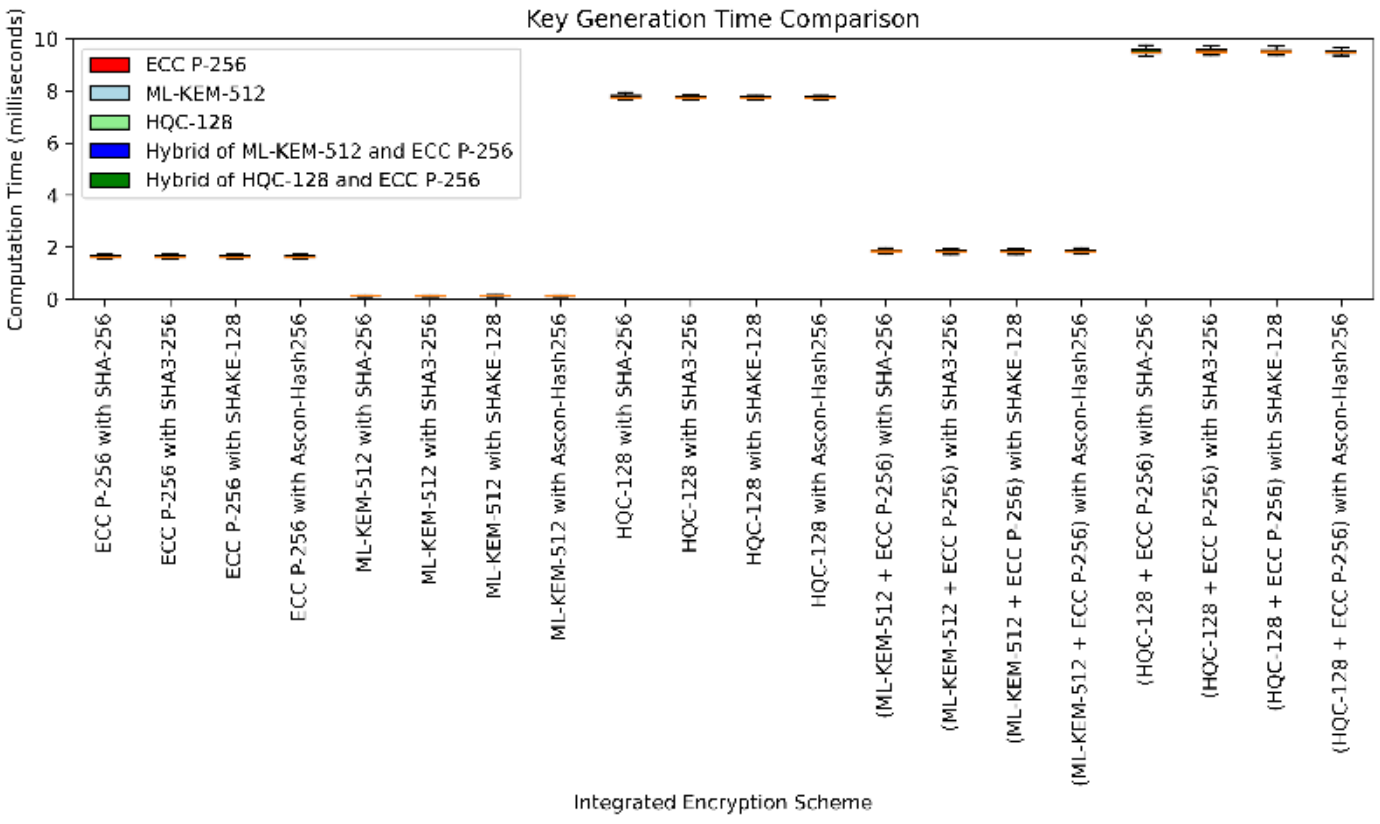

**Fig. 6.** The comparison of key generation time (unit: ms).

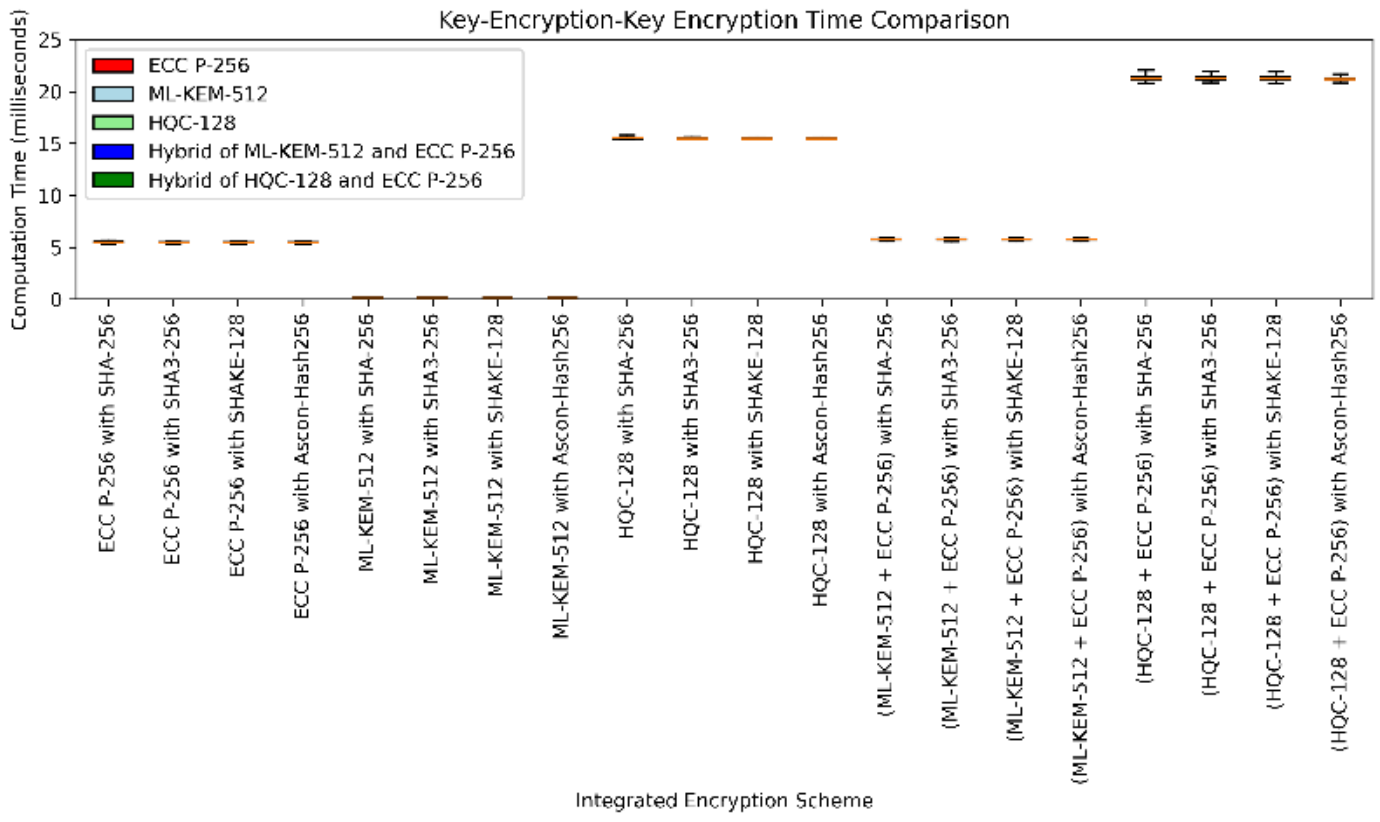

**Fig. 7.** The comparison of KEK encryption time (unit: ms).

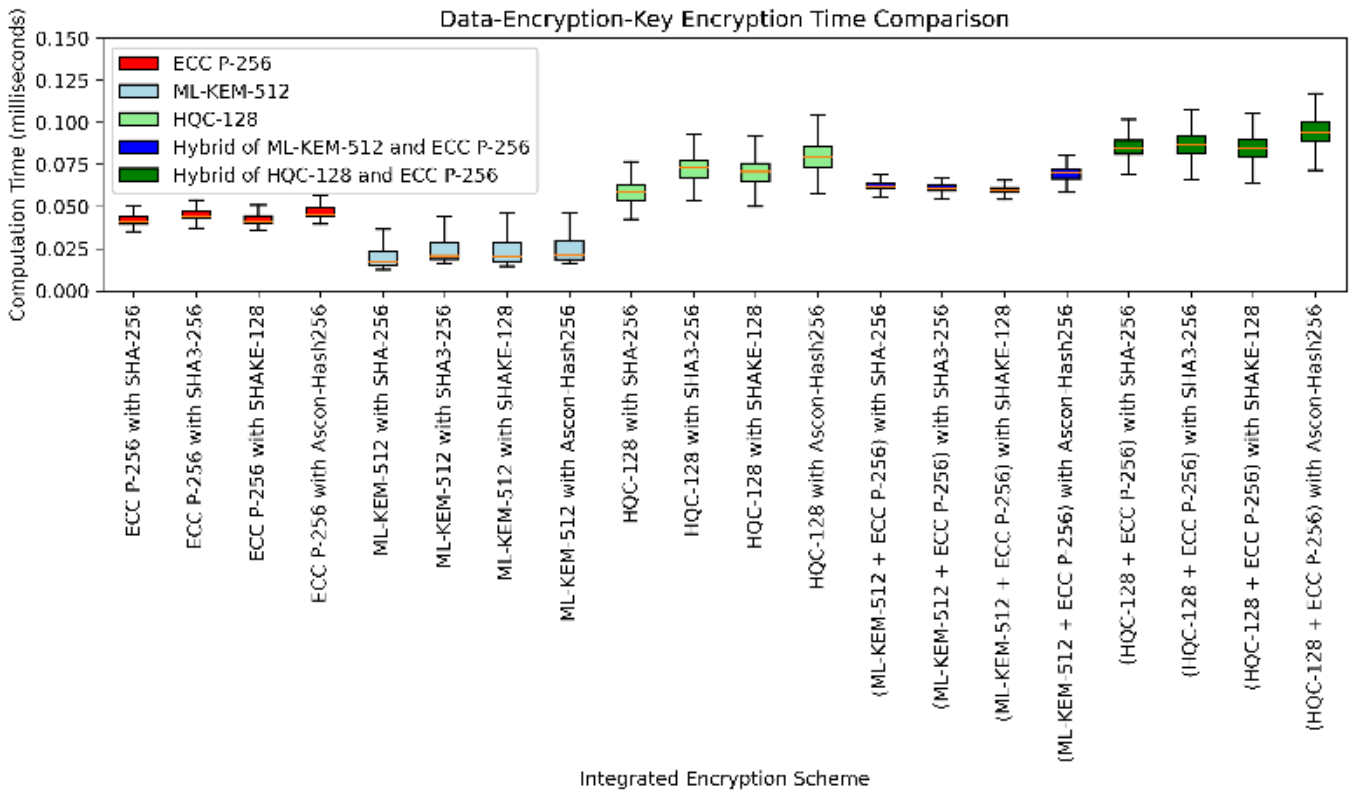

**Fig. 8.** The comparison of DEK encryption time (unit: ms).

For data encryption, the Ascon series implementations in this study employed Ascon-AEAD128, while other implementations used AES-128. As shown in Fig. 9, encryption using Ascon-AEAD128 requires less computation time than AES-128, demonstrating higher computational efficiency.

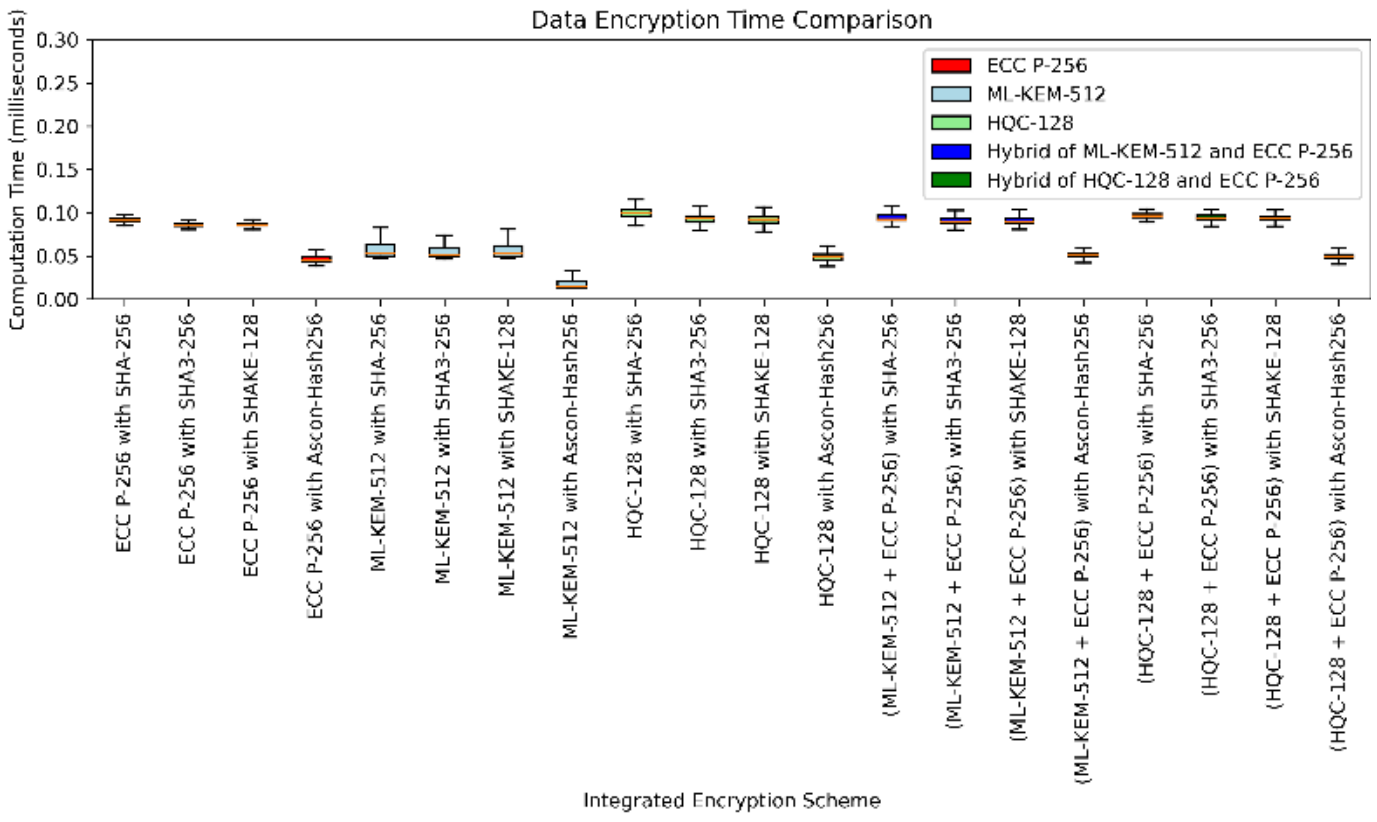

**Fig. 9.** The comparison of data encryption time (unit: ms).

Due to page limitations, only the comparison of encryption computation times is presented; however, the results for decryption computation times are consistent with those observed for encryption.

### *C. SPDU Size Comparison*

This section discusses the sizes of the EncryptedDataEncryptionKey required by each integrated encryption method when applied to V2X communications. The comparison results are summarized in Table I. ECIES requires the shortest total size because the encrypted KEK based on ECC is the shortest. Among KEM-IESs, ML-KEM produces a shorter encapsulated key value than HQC, resulting in the second-shortest total size. In the hybrid IES, the encrypted KEK size is the sum of the ECC public key and the encapsulated key value, making its EncryptedDataEncryptionKey longer than those of the other schemes. Although ECIES achieves the shortest SPDU length, it does not provide quantum resistance. Conversely, while KEM-IES offers quantum security, their main drawback is the comparatively large length of the encapsulated key values.

TABLE I

THE SIZE COMPARISON OF EncryptedDataEncryptionKey

| IES | Quantum Safe | Size of encrypted KEK | Size of *c* | Size of *t* | Total |
|---|---|---|---|---|---|
| ECIES Based on P-256 [2], [3], [4] | | 33 | 16 | 32 | 81 |
| KEM-IES Based on ML-KEM-512 | x | 768 | 16 | 32 | 816 |
| KEM-IES Based on HQC-128 | x | 4433 | 16 | 32 | 4481 |
| Hybrid IES Based on P-256 and ML-KEM-512 | x | 801 | 16 | 32 | 849 |
| Hybrid IES Based on P-256 and HQC-128 | x | 4466 | 16 | 32 | 4514 |

## V. CONCLUSIONS AND FUTURE WORK

The KEM-IES proposed in this study employs a PQC-based KEM during the key exchange phase, enhancing resistance to quantum computing attacks. Furthermore, Ascon-AEAD128 is utilized for symmetric encryption, further improving computational efficiency.

Future research could focus on designing PQC-based KEMs with shorter encapsulated key values to reduce the size of the SPDU, achieving both quantum security and suitability for transmission in V2X communications.